# Molecular Regulation of Histamine Synthesis

Hua Huang[1,2*], Yapeng Li[1], Jinyi Liang[1,3], Fred D. Finkelman[4,5]

[1]Hua Huang, the Department of Biomedical Research, National Jewish Health, Denver, CO, USA
[2]the Department of Immunology and Microbiology, University of Colorado School of Medicine, Aurora, CO, USA
[1]Yapeng Li, Department of Biomedical Research, National Jewish Health, Denver, CO, USA
[1]Jinyi Liang, Department of Biomedical Research, National Jewish Health, Denver, CO, USA
[3]Jinyi Liang, Department of Parasitology, Zhongshan School of Medicine, Sun Yat-sen University, Guangzhou, China
[4]Fred D. Finkelman, the Division of Immunobiology, Cincinnati Children's Hospital Medical Center,
[5]The Division of Immunology, Allergy and Rheumatology, Department of Medicine, University of Cincinnati College of Medicine, Cincinnati, Ohio, USA.

Correspondence:
Dr. Hua Huang, Department of Biomedical Research, National Jewish Health, 1400 Jackson Street, Denver, CO 80206. E-mail address: huangh@njhealth.org

Keywords: Histamine, Histidine decarboxylase, enhancers, promoter, gene regulation

ABSTRACT

Histamine is a critical mediator of IgE/mast cell-mediated anaphylaxis, a neurotransmitter and a regulator of gastric acid secretion. Histamine is a monoamine synthesized from the amino acid histidine through a reaction catalyzed by the enzyme histidine decarboxylase (HDC), which removes carboxyl group from histidine. Despite the importance of histamine, transcriptional regulation of *HDC* gene expression in mammals is still poorly understood. In this Review, we focus on discussing advances in the understanding of molecular regulation of mammalian histamine synthesis.

# INTRODUCTION

Bill Paul's impact on immunology is broad and enormous. Like many of his former trainees, I had the good fortune to learn from him. Bill's mentorship has nurtured my lifelong interest in studying type 2 immune responses that cause allergic diseases and protect against parasitic infections. In the early years of my laboratory, we had investigated how naïve $CD4^+$ T cells commit into T helper type 1 (Th1) cells by silencing the potential to transcribe the *Il4* gene (1-3). More recently, we extended our efforts to understand how a bi-potential basophil and mast cell progenitor acquires the capacity to transcribe a set of basophil-specific or mast cell-specific genes while simultaneously repressing transcription of a gene set that is specific for the other cell type (4). With a newly gained understanding of a network of transcription factors and their targeted enhancers (5), our laboratory has chosen to investigate the *Hdc* gene (encode histidine decarboxylase, a rate-limiting enzyme for histamine synthesis) in greater detail.

Anaphylaxis is a serious allergic reaction that is rapid in onset and includes signs and symptoms that involve the skin, gastrointestinal track, respiratory system and cardiovascular system (6). The most severe form of anaphylaxis is anaphylactic shock, which is characterized by hypotension and can cause death (6). Anaphylaxis can be caused by allergy to foods, insect venoms, medications and other agents (6). The incidence of food-induced anaphylaxis has risen at an alarming rate, especially in children, in developed countries during the past several decades and continues to rise (7-9). The economic burden of food allergy is staggering with an estimated cost of 25 billion dollars annually in the US alone (10).

Histamine plays an essential role in IgE-medicated anaphylaxis, the most common type of anaphylaxis (11-14). Histamine was first purified from ergot fungi (15) in 1910 and from human tissues (16) in 1927. Histamine has pleiotropic effects on skin and the cardiovascular, respiratory, digestive, central nervous and immune systems (17). It is a profound vasodilator that increases blood vessel permeability, allowing blood leukocytes to enter tissues to promote inflammatory responses. Relatively large quantities of histamine can cause a rapid decrease in body temperature due to massive leakage of blood plasma into the extravascular space. Rapid release of large amounts of histamine leads to anaphylaxis (12, 14). Histamine belongs to a family of biogenic amines that includes neurotransmitters, such as serotonin and dopamine, and hormones, such as epinephrine. Biogenic amines that contain one or more amine groups are formed mainly by decarboxylation of amino acids. Histamine is a monoamine synthesized from the amino acid histidine through a reaction catalyzed by the enzyme histidine decarboxylase (HDC), which removes carboxyl group from histidine (**Figure 1**). Although histamine can be synthesized by bacteria found in contaminated food (18) and in the gut of asthma patients (17, 19), in this Review, we focus on discussing advances in the understanding of molecular regulation of mammalian histamine synthesis.

**HISTAMINE-PRODUCING CELLS IN MAMMALS AND STIMULI THAT TRIGGER HISTAMINE RELEASE**

Histamine is synthesized primarily by mast cells, basophils, histaminergic neurons in the basal ganglia of the brain and enterochromaffin-like cells (ECL) in the stomach. These cells produce large amounts of histamine and are thought to be the major histamine-producing cells (**Figure 2**). They continuously synthesize histamine, which is then linked to the carboxyl group of heparin and stored in intracellular granules until the cells receive the appropriate activating stimulus. Upon external stimulation, these cells degranulate, releasing the stored histamine. Stimuli that trigger histamine release by these major histamine-producing cells have been reviewed extensively (20-25). Antigen crosslinking of antigen-specific IgE bound to the high affinity IgE receptor, FcεRI, on the mast cell and basophil surface is the most robust stimulus that triggers histamine release by these cells (20-23). Substance P and allergy-inducing drugs that bind to G-protein-coupled receptors can also trigger basophils and mast cells to release histamine via different signaling pathway (23, 26). Additionally, complement components, such as the C3a and C5a "anaphylatoxins," have also been shown to induce histamine release by mast cells (27). Many cytokines, including IL-3, IL-18, IL-33, GM-CSF and SCF, promote histamine synthesis (28-30). In general, cytokines alone do not induce histamine release although it remains controversial whether IL-33 can have this effect. Some reports describe that IL-33 stimulates histamine release (31, 32), while other reports dispute this (33, 34). It is suggested that IL-33 alone does not induce histamine release by basophils, but enhances histamine release in response to IgE/FcεRI crosslinking (35) (**Figure 2**).

Additional histamine-producing cells have also been identified, including T cells (36), dendritic cells (37), macrophages (38, 39) and epithelial cells (40, 41) (**Figure 2**). In contrast to mast cells and basophils, these cells produce relative small quantities of histamine and do not store it in their cytoplasm (42). The small amounts of histamine that they produced are released without external stimulation (42). The biological significance of the small amounts of histamine produced by these minor histamine-producing cells remains unclear. Cell type-specific deletion of the *Hdc* gene, which encodes histidine decarboxylase, an enzyme essential for histamine synthesis, would shed light on the role of histamine synthesis and secretion by the minor histamine-producing cells.

**HISTIDINE DECARBOXYLASE AND HISTAMINE SYNTHESIS IN MAMMALS**

After several groups purified mammalian HDC protein from fetal rat liver and mouse mastocytoma P-815 cells (43-45), a cDNA that encodes this protein was subsequently cloned (46, 47). The *Hdc* gene encodes HDC protein, which has a molecular mass of 74 kDa and is a proenzyme with little or no enzyme activity. Once the proenzyme is cleaved at the site near its c-terminus, presumably by Caspase-9, it yields a 53 kDa N-terminal and a 20 kDa C-terminal subunit. The 20 kDa C-terminal subunit is believed to possess inhibitory activity (48). The 53 kDa N-terminal subunit forms a homodimer that is an active decarboxylase (48, 49). HDC is the primary enzyme that catalyzes histamine synthesis. Mice deficient in the *Hdc* gene fail to synthesize histamine and have reduced or absent IgE-mediated anaphylactic responses (50-53). Several potent HDC inhibitors have been identified, including the histidine derivatives α-fluoromethyl histidine, histidine methyl ester and pirodoxal histidine methyl ester (54-56). However, these HDC inhibitors have not been further developed for clinical use.

**HDC GENE EXPRESSION AND HISTAMINE SYNTHESIS IN BASOPHILS AND MAST CELLS**

*Hdc* gene expression and histamine synthesis are regulated both positively and negatively by a range of factors. Notably, crosslinking of FcεRI by antigen binding to FcεRI-associated IgE increases mast cell *Hdc* mRNA expression and histamine synthesis (57, 58). These mast cell activation-induced increases in *Hdc* mRNA expression and histamine synthesis are also induced by phorbol 12-myristate 13-acetate (59). *Hdc* mRNA expression and histamine synthesis also increase as immature mast cells undergo maturation. Bone marrow-derived mast cells (BMMCs) appear immature because they contain relatively little histamine and express relatively low levels of FcεRI (60). These immature mast cells develop into mature mast cells with higher amounts of histamine in vivo if they are adoptively transferred into the peritoneal cavity (61). However, it is not clear if in vivo exposure to IgE promotes maturation and increases *Hdc* mRNA expression.

In this regard, we demonstrated that chlorotoxin, which induces mast cell maturation (62), strongly upregulates *Hdc* gene expression in BMMCs within few hours after the treatment (5). The mechanism by which chlorotoxin enhances *Hdc* gene transcription remains to be determined. It is conceivable that chlorotoxin activates mast cells by binding to an acidic glycosphingolipid, ganglioside G, that has been shown to be expressed on the mast cell surface (62). Chlorotoxin-triggered signals in mast cells then activate transcription factors that directly and rapidly promote *Hdc* gene transcription. It is unknown whether bacteria in the gut of allergic patients can promote *Hdc* mRNA and histamine synthesis by producing substances similar to chlorotoxin.

In line with the notion that factors promoting mast cell maturation also enhance histamine synthesis, cytokines that promote basophil and mast cell maturation, such as IL-3, IL-18, IL-33, GM-CSF and SCF, have also been reported to increase HDC activity (28-30, 63). It is unclear whether these cytokines regulate *Hdc* gene transcription by increasing the expression of the genes that produce *Hdc* gene-activating transcription factors or by activating already produced transcription factors to induce transcription of the *Hdc* gene. Other substances, including chemokines, neuropeptide substance P, and IL-1α have also been reported to induce *Hdc* mRNA and histamine synthesis (64, 65).

In contrast, mitochondrial uncoupling protein 2, a mitochondrial transporter protein that transfers anions from the inner to the outer mitochondrial membrane and protons from the outer to the inner mitochondrial membrane, inhibits *Hdc* mRNA expression and histamine synthesis, possibly by suppressing the production of reactive oxygen species (66). Substances found in fruits and vegetables, such as quercetin (67), and in green tea, such as epigallocatechin gallate, also potently inhibit HDC (68). More detailed examination of negative regulators of *Hdc* mRNA expression should promote development of agents that may be able to prevent and treat food allergy and other histamine-mediated allergic inflammatory disorders.

The human *HDC* gene is located in the 15q21.2 region of chromosome 15. It contains 12 exons (69) (**Figure 3**). Eight predicted isoforms can be generated by alternative splicing and two actual isoforms have been described (70). *HDC* mRNA is expressed broadly in many organs, with the highest expression levels found in the gallbladder, stomach and lung (71). Because the RNA-seq data for normal tissues in the Human Protein Atlas were obtained from intact tissues, it is not clear whether the human *HDC* gene is expressed predominantly in known histamine-producing cells, such as mast cells and enterochromaffin-like cells (ECL) in high *HDC*-expressing tissues, or predominantly in other cell types in those tissues. In contrast to the human *Hdc* gene, the mouse *HDC* gene is located in chromosome 2 (72). It resembles the human gene in that it contains 12 exons, is expressed broadly in many tissues with the highest expression levels in lung, ovary and subcutaneous fat pads (72, 73) and is 86% homologous with the human gene (https://www.ncbi.nlm.nih.gov/homologene/20490); however, there are only 3 predicted isoforms and no isoform, other than the classical one, have been found for murine *Hdc* (72).

There is still limited knowledge of how *Hdc* gene expression is regulated transcriptionally. Most previous work has concentrated on the promoter region of this gene. Deletion analysis of *Hdc* promoter-driven luciferase reporter gene transcription demonstrated that the transcription factor SP1 binds to a GC box (GGGGCGGGG) found in both the human and mouse *Hdc* gene promoters (72, 74). Several promoter elements have been reported to negatively regulate *Hdc* gene transcription. For example, the transcription factors YY1 and KLF4 have been shown to negatively regulate the *Hdc* gene by suppressing SP1 in a gastric cancer cell line (75, 76).

In contrast, *Hdc* gene expression is positively regulated by the transcription factor GATA binding protein 2 (GATA2), a member of the GATA family of transcription factors. GATA2 is critical for survival and proliferation of hematopoietic stem cells (77, 78), granulocyte-monocyte progenitor differentiation (79), and basophil and mast cell differentiation (80, 81) and is required for connective tissue mast cell development (5). In contrast, basophil development is not affected in connective tissue-specific *Gata2*-deficient mice (5). We have also found that mucosal and connective tissue-specific *Gata2*-deficient mice fail to develop both mucosal and connective tissue mast cells, indicating that GATA2 is required for both mucosal and connective tissue mast cell development (Li et al., unpublished data). To distinguish the role of GATA2 in regulating the *Hdc* gene from its role in mast cell development, we used an inducible gene deletion method to delete the *Gata2* gene from mast cells after they had fully differentiated. In this inducible gene deletion model, the enzyme Cre is fused to the estrogen receptor (ER) and the ER-Cre fusion product is induced to enter the cell nucleus to cleave a floxed gene of interest by the ER ligand 4-hydroxytamoxifen (82). Using this method, we demonstrated that GATA2 plays a critical role in regulating *Hdc* gene expression in even fully differentiated mast cells. However, in contrast to its role in mast cell development, GATA2 is not needed for survival of fully differentiated mast cells (83).

More recently, our group has used active histone mark ChIP and reporter gene transcription assays to identify and characterize two *Hdc* enhancers in mast cells. Epigenomic studies demonstrate that monomethylation of lysine 4 on histone 3 (H3K4me1) marks genes that are poised to be transcribed, whereas acetylation of lysine 27 on histone 3 (H3K27ac) identifies genes that are actively being transcribed. The combined presence of H3K4me1 and H3K27ac modifications predicts enhancer activity (84-88). Our H3K4me1 and H3K27ac ChIP-seq analysis of BMMCs identified two putative *Hdc* enhancers located -8.8 kb upstream and +0.3 kb downstream from the transcription start site (TSS) of the *Hdc* gene (**Figure 1**). We demonstrated that the -8.8 kb *Hdc* enhancer, but not the +0.3 kb *Hdc* enhancer, increases minimal *Hdc* promoter activity in a luciferase reporter gene transcription assay. The transcription factor MITF binds to the -8.8 *Hdc* enhancer and drives its enhancer activity. Indeed, MITF overexpression largely restores *Hdc* gene expression in *Gata2*-deficient mast cells. Our study also suggests that GATA2 induces MITF and that these two transcription factors together direct full *Hdc* gene transcription in mast cells in a feed-forward manner. However, it is not certain that the -8.8 kb *Hdc* enhancer is fully responsible for positive regulation of the *Hdc* gene, because in vivo importance of the +0.3 kb *Hdc* enhancer in *Hdc* gene transcription cannot be ruled out by the luciferase reporter gene transcription assay alone (5).

Despite remarkable progress in genome-wide annotation of potential enhancers, functional validation of annotated enhancers remains an unmet challenge. Transgenic mice, reporter gene assay, and CRISPR/Cas9 genome editing have been used to validate the biological functions of enhancers identified by histone marks. Each of these methods has its strengths and weaknesses (89, 90). The reporter gene assay has been widely used to assess enhancer activity. It is simple, rapid and efficient at assessing promoter and enhancer activity in transiently or stably transfected cell lines. The limitation of the transient reporter gene assay is that it does not measure promoter and enhancer activity in the context of chromatin. Despite this disadvantage, this reductionist approach is useful for assessing binding of transcription factors to *cis* regulatory elements in accessible regions. It has been reported that ~60% of annotated enhancers show enhancer activity by the luciferase reporter gene assay (86, 91-94). The in vivo function of the -8.8 *Hdc* enhancer requires further investigation.

**HISTAMINE SYNTHESIS IN THE CENTRAL NERVOUS SYSTEM AND THE STOMACH**

In addition to its activity as a vasoactive mediation, histamine is a neurotransmitter and a regulator of gastric acid secretion. HDC mRNA is expressed in the brain exclusively in the basal ganglia (95). Specific ablation of histaminergic neurons leads to repetitive movements (96), that resemble the signs of Tourette syndrome (97). Consistent with this, a nonsense mutation at the human HDC gene (W317X) has been identified in a family of patients with this syndrome (97, 98) and mice completely deficient in *Hdc* gene transcription develop a Tourette-like syndrome (97, 99). However, the mechanisms involved in *Hdc* gene regulation in the basal ganglia are currently unknown. In the stomach, histamine is synthesized in ECL and is released from these cells upon gastrin and acetylcholine stimulation. The released histamine then stimulates parietal cells to secrete stomach acid (25, 100). Mice deficient in the *Hdc* gene fail to fully acidify their gastric contents (100), which can lead to indigestion, diarrhea, constipation or rectal itching (101). Clinically, histamine 2 ($H_2$) receptor antagonists, such as ranitidine, are currently used to ameliorate stomach hyperacidity and peptic ulcer disease by blocking this receptor on the hydrochloric acid-producing parietal cells in the stomach (Ranitidine: a review of its pharmacology and therapeutic use in peptic ulcer disease and other allied diseases. Brogden RN, Carmine AA, Heel RC, Speight TM, Avery GS. Drugs. 1982 Oct;24(4):267-303.). At present, it is not known how the *Hdc* gene is regulated in ECL cells. It is most likely that different transcription factors are used to regulate the *Hdc* gene in basal ganglia and ECL cells.

**CONCLUDING REMARKS**

HDC is the rate-limiting enzyme for histamine synthesis. Understanding transcriptional regulation of the *Hdc* gene will advance our knowledge about how this gene detects extracellular stimuli and increases its transcription, leading to histamine synthesis, replenishment and accumulation that exacerbate allergic inflammation and anaphylaxis. Fine mapping of critical transcription factors and their authentic binding sites within the *Hdc* promoter and enhancers should promote identification of regulatory variants that influence allergy susceptibility and severity. Today, Bill Paul's teaching and his large body of work on IL-4 continues to inspire our fascination with type 2 immunity.

**ACKNOWLEDGMENTS**

Research reported in this article was supported by grants from the National Institutes of Health R01AI107022 and RO1AI083986 (H.H.), R01AI113162 (F. F.), and a fund provided by Sun Yet-Sen University (J.L.).

**FIGURE 1 |** Histamine synthesis

**FIGURE 2 |** Histamine-producing cells and stimuli that trigger histamine release.

**FIGURE 3 |** Genomic structures of the human and mouse *HDC* gene.

Red bars indicate the enhancers we described.

**Figure 1**

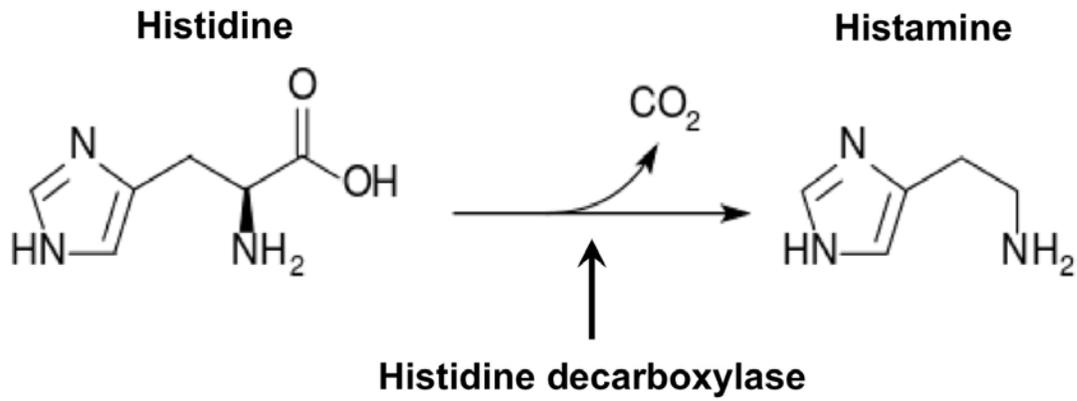

**Figure 2**

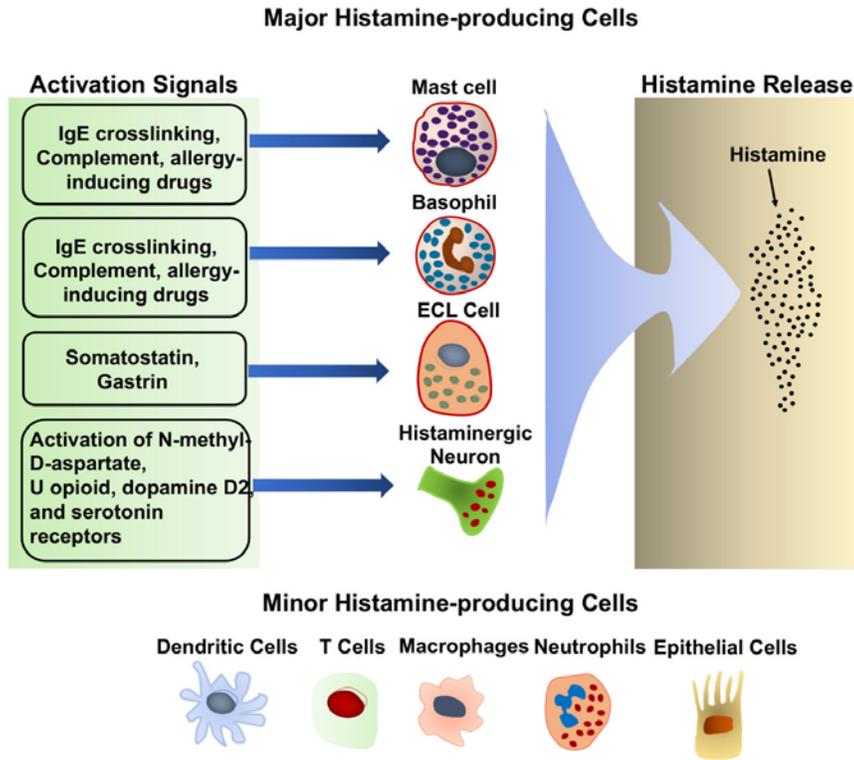

Figure 3

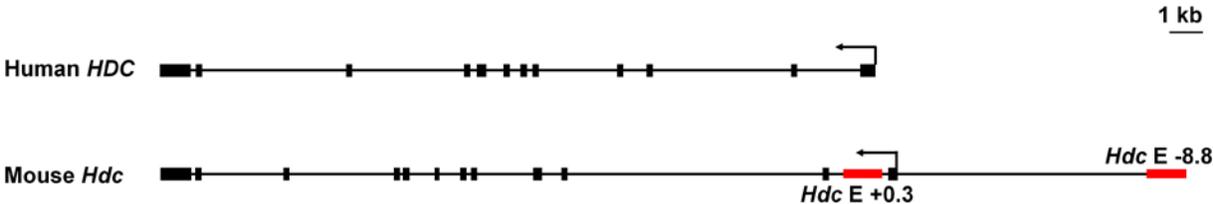